\documentclass{article}
\usepackage{arxiv}
\usepackage{subcaption}
\usepackage{times}
\usepackage{epsfig}
\usepackage{graphicx}
\usepackage{amsmath}
\usepackage{amssymb}
\usepackage{booktabs}
\usepackage{url}
\usepackage{hyperref}
\usepackage{enumitem}
\usepackage{float}

\def\BibTeX{{\rm B\kern-.05em{\sc i\kern-.025em b}\kern-.08em
    T\kern-.1667em\lower.7ex\hbox{E}\kern-.125emX}}
\usepackage{balance}

\begin{document}

\title{Probabilistic Modeling of Inter- and Intra-observer Variability in Medical Image Segmentation}

\author{ 
{Arne Schmidt} \\
Department of Computer Science and Artificial Intelligence\\ 
University of Granada\\ 
Granada, Spain \\
\And
{Pablo Morales-Álvarez} \\
Department of Statistics and Operations Research\\
University of Granada\\ 
Granada, Spain \\
\And
{Rafael Molina}\\
Department of Computer Science and Artificial Intelligence\\ 
University of Granada\\ 
Granada, Spain \\
\thanks{This work has received funding from the European Union’s Horizon 2020 research and innovation programme under the Marie Skłodowska Curie grant agreement No 860627 (CLARIFY Project), from the Spanish  Ministry  of  Science  and Innovation under project PID2019-105142RB-C22, and by FEDER/Junta de Andalucía-Consejería de Transformación Económica, Industria, Conocimiento y Universidades under the project P20\_00286.}
}
\renewcommand{\shorttitle}{Probabilistic Modeling of Inter- and Intra-observer Variability in Med. Im. Seg.}
\maketitle
\begin{abstract}
   Medical image segmentation is a challenging task, particularly due to inter- and intra-observer variability, even between medical experts. In this paper, we propose a novel model, called Probabilistic Inter-Observer and iNtra-Observer variation NetwOrk (Pionono). It captures the labeling behavior of each rater with a multidimensional probability distribution and integrates this information with the feature maps of the image to produce probabilistic segmentation predictions. The model is optimized by variational inference and can be trained end-to-end. It outperforms state-of-the-art models such as STAPLE, Probabilistic U-Net, and models based on confusion matrices. Additionally, Pionono predicts multiple coherent segmentation maps that mimic the rater's expert opinion, which provides additional valuable information for the diagnostic process. Experiments on real-world cancer segmentation datasets demonstrate the high accuracy and efficiency of Pionono, making it a powerful tool for medical image analysis.
\end{abstract}
%%%%%%%%% BODY TEXT
\section{Introduction}
% Introducing the problem: AI algorithms can help in diagnostic processes and have shown impressive performances. But often there is a high variability even in expert opinions.
Artificial Intelligence (AI) algorithms have shown remarkable progress in image analysis, holding great promise for faster and more accurate diagnostic procedures \cite{kohl_probabilistic_2018, schmidt_efficient_2022, schmidt_probabilistic_2023,  amgad_structured_2019}. Nevertheless, in medical practice, there exists a high degree of variability among the opinions of different medical experts, even when the same expert assesses the same data at different times. This inter- and intra-observer variability has been reported across various tasks, including MRI-based segmentation of HCC lesions \cite{covert_intra-_2022}, lung cancer segmentation in CT scans \cite{kulberg_inter-observer_2021}, and multiple fields in pathology \cite{mahbod_cryonuseg_2021, allard_intraobserver_2018, brochez_inter-observer_2002, nir_automatic_2018}. It leads to uncertainties when applying AI models because in contrast to other classification tasks, there is not a single ground truth.

% Bayesian neural networks have addressed uncertainty
Especially in the medical domain, the careful modeling of uncertainties in its different forms has a high priority to minimize the risk of relying on incorrect predictions \cite{linmans_predictive_2023, schmidt_probabilistic_2023, kwon_uncertainty_2020, kohl_probabilistic_2018, kandemir_variational_2016}. In recent years, probabilistic methods, such as Bayesian Neural Networks \cite{gal_dropout_2016} and sparse Gaussian processes \cite{kandemir_variational_2016, wu_combining_2021} have gained more and more attention, because they are able to account for uncertainties in a sound manner. They showed promising results when modeling uncertainty in the network weights \cite{gal_dropout_2016}, data ambiguities\cite{kwon_uncertainty_2020} or attention weights \cite{schmidt_probabilistic_2023}. 
\begin{figure}[t]
\begin{center}
\includegraphics[width=0.5\linewidth]{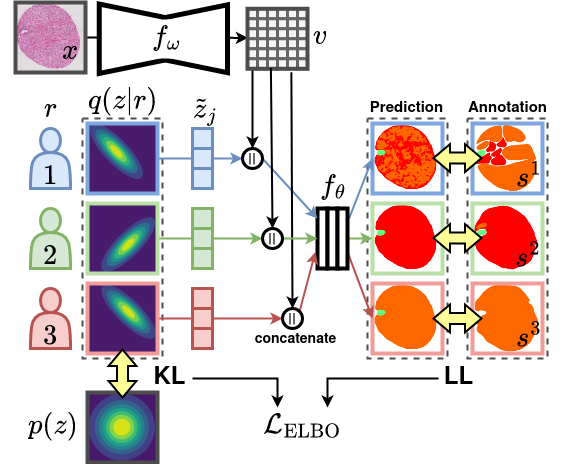}
\end{center}
   \caption{The proposed Pionono model. The labeling behaviour of each rater $r$ is represented by a multivariate Gaussian distribution $q(z|r)$. The drawn samples $\tilde{z}_j$ are concatenated with the extracted features $v$ of $f_\omega$ and then fed into the segmentation head $f_\theta$. The output simulates the inter- and intra-observer variability of annotations and is optimized using the real annotations $s^r$ of each rater. The model is trained end-to-end with a combination of log-likelihood loss (LL) and Kulback Leibler (KL) divergence between posterior and prior, combined in the overall loss $\mathcal{L}_\text{ELBO}$.}
\label{fig:model}
\end{figure}
% Motivating our algorithm: We have developed a probabilistic deep learning model that explicitly models this inter- and intra observer variability. This leads to an improved performance and allows to simulate different expert opinions for a given test image.
Although inter- and intra-observer variability is often mentioned as a key challenge when applying AI to medical data \cite{mahbod_cryonuseg_2021, li_em-based_2018, silva-rodriguez_going_2020, lopez-perez_learning_2021}, to the best of our knowledge there is no method that explicitly models these two types of uncertainty for medical image segmentation.

To address this gap, we propose a novel approach called the Probabilistic Inter-Observer and iNtra-Observer variation NetwOrk (Pionono), depicted in Figure \ref{fig:model}. This model accurately accounts for inter- and intra-observer variability using probabilistic deep learning. Specifically, each rater's labeling behavior is represented as a probability distribution in latent space, and optimized using the log Evidence Lower BOund (ELBO) in an end-to-end training process. The variance of each rater's distribution models the intra-observer variability, while the differences between the distributions models the inter-observer variability. When two raters exhibit similar labeling behavior, their probability distributions overlap substantially, while different labeling behavior results in a small overlap of distributions. 

The approach is validated in extensive experiments of prostate and breast cancer segmentation, using 'gold' labels. They reflect the expert agreement to show that our probabilistic modeling improves the predictive performance and estimates the predictive uncertainty. Furthermore, we also test its capability to model each rater's labeling behavior. As shown in the experiments, it can simulate expert opinions for a given test image in a consistent manner, providing a realistic estimation of “what expert X would say in this case". 
Our contributions can be summarized as follows:
\setlist{nolistsep}
\begin{itemize}[noitemsep]
    \item We propose Pionono, a probabilistic deep learning model that uses probability distributions in latent space to represent inter- and intra-observer variability. It can be trained with labels of multiple raters.
    \item The model is able to provide accurate segmentation predictions (compared to the expert agreement and different expert opinions), outperforming existing state-of-the-art algorithms such as STAPLE, Probabilistic U-Net and models based on global or local confusion matrices.
    \item Pionono provides uncertainty estimations that indicate areas where the predictions are not conclusive.
    \item The proposed model can provide several coherent segmentation hypotheses, simulating different medical experts.
\end{itemize}
% Related Work

% Overview of Pros and Cons of each method
\section{Related Work} \label{sec:related_work}
In this section, we review existing methods of probabilistic deep learning and crowdsourcing for medical images and highlight the differences to our model.

\textbf{Probabilistic Deep Learning.}
% Describe existing probabilistic approaches and different uncertainties to be estimated
As already indicated, probabilistic approaches such as Bayesian neural networks \cite{gal_dropout_2016, abdullah_review_2022, kwon_uncertainty_2020, linmans_predictive_2023} and sparse Gaussian processes \cite{kandemir_variational_2016, wu_combining_2021, schmidt_probabilistic_2023} have shown promising results in a multitude of tasks in the medical image domain, modeling different sources of uncertainties. Often, a general predictive uncertainty is addressed using probabilistic weight parameters \cite{abdullah_review_2022}. This uncertainty can be bisected into model and data uncertainty which originate from model parameters or data ambiguities, respectively \cite{kwon_uncertainty_2020}. Other approaches have modeled the uncertainty of missing instance labels in multiple instance learning \cite{lopez-perez_deep_2022, schmidt_probabilistic_2023} or uncertainty of out of distribution samples \cite{linmans_predictive_2023}. 
% Highlight differences to Prob-Unet
The uncertainty in annotations has previously been addressed by the Probabilistic U-Net \cite{kohl_probabilistic_2018} (\textit{Prob U-Net}), which encodes the labeling behavior in a latent random variable. The model is trained as a variational autoencoder with an encoder network predicting the latent distribution.
This approach models a general variability in annotations but lacks the explicit modeling of inter- and intra-observer variability. Therefore, it is not able to incorporate the rater information during training and cannot simulate expert opinions. 

\begin{table}
\small
\setlength{\tabcolsep}{1pt}
\begin{center}
\begin{tabular}{rrcrcrcrc}
\toprule
       & (i) & Prob.  & (ii) & Coh. & (iii) & Exp.  & (iv) &  Scale \\
Method & &  Uncert. & &  Segm. &&  Opinion & & \\
\midrule
STAPLE  
& & \includegraphics[width=0.35cm]{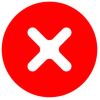} 
& & \includegraphics[width=0.35cm]{images/cross.png} 
& & \includegraphics[width=0.35cm]{images/cross.png} 
& & \includegraphics[width=0.35cm]{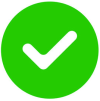}     \\
Prob U-Net  
& & \includegraphics[width=0.35cm]{images/check.png} 
& & \includegraphics[width=0.35cm]{images/check.png} 
& & \includegraphics[width=0.35cm]{images/cross.png} 
& & \includegraphics[width=0.35cm]{images/check.png}     \\
CM global 
& & \includegraphics[width=0.35cm]{images/cross.png} 
& & \includegraphics[width=0.35cm]{images/cross.png} 
& & \includegraphics[width=0.35cm]{images/check.png} 
& & \includegraphics[width=0.35cm]{images/check.png}     \\
CM pixel 
& & \includegraphics[width=0.35cm]{images/cross.png} 
& & \includegraphics[width=0.35cm]{images/cross.png} 
& & \includegraphics[width=0.35cm]{images/check.png} 
& & \includegraphics[width=0.35cm]{images/cross.png}     \\
Pionono  
& & \includegraphics[width=0.35cm]{images/check.png} 
& & \includegraphics[width=0.35cm]{images/check.png} 
& & \includegraphics[width=0.35cm]{images/check.png} 
& & \includegraphics[width=0.35cm]{images/check.png}     \\
\bottomrule
\end{tabular}
\end{center}
\caption{For AI segmentation models to achieve the best possible diagnostic support, they should address four key issues: (i) provide a \textit{probabilistic uncertainty} estimation, not only a single prediction for a test image; (ii) provide multiple \textit{coherent segmentation} hypotheses; (iii) simulate different \textit{expert opinions} for better explainability and decision support; (iv) \textit{scale} to a higher amount of raters in the case that more data from different hospitals can be integrated.} \label{tab:related_work}
\end{table}

\textbf{Crowdsourcing.}
While existing crowdsourcing methods aim to capture inter-observer variability in the training labels, this variability is often not reflected in the test predictions by probabilistic outputs \cite{kohl_probabilistic_2018}. The intra-observer variability is often not modeled at all, although it is often mentioned as a challenge in literature \cite{mahbod_cryonuseg_2021, li_em-based_2018, silva-rodriguez_going_2020, allard_intraobserver_2018, covert_intra-_2022}.

% Majority voting/STAPLE
One way to handle multiple annotations is label fusion. With this method, the annotations of different raters are merged to a single set of labels. The “Simultaneous Truth and Performance Level Estimation" (\textit{STAPLE}) mechanism performs label fusion with a probabilistic estimate of the true labels by weighting each segmentation depending upon the estimated performance level of each rater \cite{warfield_simultaneous_2004}. A supervised network can then be trained on these fused labels.
% Confusion Matrices
More dedicated approaches incorporate different rater labels using confusion matrices (CM), for example for classification of image patches with Gaussian processes \cite{lopez-perez_learning_2021} or image segmentation with global confusion matrices \cite{tanno_learning_2019} (\textit{CM global}). 
% Pixel Confusion Matrix
In this direction, also pixel-wise confusion matrices were explored for semantic segmentation that are estimated by a dedicated deep neural network \cite{zhang_disentangling_2020} (\textit{CM pixel}). These models have shown promising results, but come with a conceptual problem: They assume that the pixels are statistically independent of each other, although neighboring pixels have a high correlation. Therefore, the output of different segmentation hypotheses is not coherent. Furthermore, the predictions of the mentioned approaches \cite{tanno_learning_2019, zhang_disentangling_2020} are not modeled by a predictive distribution, but by a deterministic point estimate. While the global confusion matrix approach \cite{tanno_learning_2019} has a limited expressiveness, the pixel-wise calculation \cite{zhang_disentangling_2020} is hard to scale for multiple raters, because for each rater, a complete deep neural network must be trained and stored.

Pionono unites the advantages of probabilistic and crowdsourcing methods. 
%It models each rater's behavior with one probability distribution, which is optimized during training. 
We summarize a comparison of different characteristics of Pionono and related methods for image segmentation in Table \ref{tab:related_work}.

\section{Methods}
In this section, we outline the background of the proposed method.
It is implemented in the Pytorch \cite{paszke_pytorch_2019} framework and is publicly available at \url{https://github.com/arneschmidt/pionono_segmentation}.

\subsection{Problem Definition} 
Let $X = \{x_i \in \mathbb{R}^{H \times W \times 3} \}_{i=1,..,N}$ be a set of images, $S^r = \{s_i^r \in \mathbb{R}^{H \times W \times C}\}_{i=1,..,N}$ the corresponding segmentation maps with image dimensions $H \times W$ and $C$ the number of classes. The segmentation maps are provided by different raters $r \in R =  \{1,2,..,M\}$. Some or all images can be segmented by multiple raters, such that some segmentation maps $s_i^r$ can be empty. The proposed model does not require any overlap of the sets of annotated images. 

If there are images available with segmentations assigned by expert agreements (so-called gold labels), the model should be able to predict a gold distribution over outputs $p(S^{\text{gold}})$ with the mean estimating the segmentation and the variance estimating the uncertainty. In any case, the model should model different segmentation hypotheses for the raters $\{p(S^r); r = 1,2,..,M\}$ for diagnostic decision support.

\subsection{Proposed Model} \label{sec:proposed_model}

% Feature extraction
% Segmentation head
% ELBO optimization
% Predictions (including gold predictions)
First, we introduce the common segmentation backbone $f_{\omega}$ with trainable weights $\omega$. We use the well-known U-Net architecture \cite{ronneberger_u-net_2015} with a Resnet34 feature extractor \cite{he_deep_2016}. This model takes an image $x_i $ and extracts a feature map $v_i \in \mathbb{R}^{H \times W \times L}$ with $H \times W$ being the image resolution and $L$ the dimensions of the feature vectors ($L=16$ in the case of U-Net).
We denote the feature extraction as
\begin{equation}
    v_i = f_{\omega}(x_i).
\end{equation}
Based on these feature vectors, we could perform segmentation with a segmentation head $f_{\theta}$:
\begin{equation}
    s_i = f_{\theta}(v_i).
\end{equation}
%Note that this classification is performed for each pixel, such that a segmentation map in the dimension of the image is predicted.

Now we extend this model to incorporate the \textit{inter- and intra-observer variation}. The segmentation maps are influenced by the rater's experience, assessment, and personal choices.
To encode the labeling behavior, we use a random vector $z \in \mathbb{R}^D$. In practice, $D=8$ are enough dimensions to reflect different labeling behaviors. 
We define a prior distribution $p(z) = \mathcal{N}(z|0, \sigma_\text{prior} * I)$ which encodes a generic labeling behavior without further information about the rater. It is possible to encode prior knowledge in this distribution, but we present a general model and leave this for future work. We set $\sigma^2_\text{prior} = 2.0$, because we observe a realistic variability in the output for this value. In section \ref{sec:ablation_studies} we prove that the model is robust for different settings of hyperparameters $D$ and $\sigma^2_\text{prior}$.

Now, the posterior distribution of $p(z|r)$ that depends on the rater $r$ should be found. We approximate it with one multivariate Gaussian distribution for each rater:
\begin{equation}
    q(z|r) = \mathcal{N}(z|\mu^r, \Sigma^r) \ \forall r=1,..,M
\end{equation}
where $\{\mu^r, \Sigma^r\}^{r=1,..,M}$ are trainable parameters.
The variance of each distribution $q(z|r)$ models the $intra-observer$ variability. The differences between the distributions for different raters model the $inter-observer$ variability.
To obtain the predictive gold distribution we add another 'rater' $r=M+1$ represented by an additional gold distribution $q$ which is trained with the available gold segmentations. During prediction, this distribution provides the estimated agreement between experts.

The segmentation head $f_{\theta}$, parametrized by weights $\theta$, must be adapted to take the random vector $z$ into account. The approximated predictive distribution is then obtained by:
\begin{equation} \label{eq:q_prediction}
    q(s_i|x_i, r, \omega, \theta) = \int f_{\theta}(v_i, z) q(z|r) d z.
\end{equation}
The closed-form calculation is not feasible and therefore we approximate it by Monte Carlo (MC) sampling:
\begin{equation}
    \tilde{s}_{i, j}|x_i, r = f_{\theta}(v_i, \tilde{z_j}) \text{; } \tilde{z_j} \sim q(z|r)
\end{equation}
with $j=1,...,K$ indexing the MC samples.
In practice, we concatenate the feature maps $v_i$ and the latent vector $\tilde{z_j}$, which is broadcasted to the image size, leading to a feature map with dimensions $H\times W \times(L+D)$. The segmentation head consists of three layers with 1x1 convolutions and 16 filters in the first two layers and $C$ filters in the last layer.

\subsection{Training}
First, all posterior distributions $q(z|r)$ are initialized randomly. Each initial value of the mean vectors $\mu^r$ is independently drawn from a distribution $\mathcal{N}(0,\sigma^2_{\text{post}})$. 
%We have tested $\sigma^2_{\text{post}}=4,8,16$ and found that  $\sigma^2_{\text{post}}=8$ shows the best results, see the ablation studies \ref{sec:ablation_studies}. 
We set $\sigma^2_{\text{post}}=8$, because this initializes the mean vectors sufficiently different for a good optimization. In section \ref{sec:ablation_studies} we show, that the model is robust to other settings of this value.
The covariance matrices $\Sigma^r$ are initialized with $\sigma_\text{prior} * I$.

To optimize the parameters $\{\mu^r, \Sigma^r\}^{r=1,..,M}$ of the probability distribution $q(z|r)$, we maximize the ELBO:
%\begin{equation}
%    KL(q(z, \omega, \theta| X, S^r)| p(z, \omega, \theta| X, S^{r})) )
%\end{equation}
%for each rater $r$.
%As in several previous works \cite{kingma_variational_2015, kandemir_variational_2016, gal_dropout_2016}, this can be achieved by maximizing the ELBO
\begin{equation} \label{eq:elbo}
    % \mathcal{L}_{ELBO} =  \mathbb{E}_{q(s|x, r, \omega, \theta)} log \ p(s|x, r) - KL(q(z|r) | p(z)).
    \mathcal{L}_{ELBO} =  \mathbb{E}_{q} log \ p(S^r|X, r, \omega, \theta) - \lambda \textit{KL}(q(Z|r) | p(Z)).
\end{equation}
with distribution $q$ as defined in eq. \ref{eq:q_prediction}. The first term defines a log-likelihood (LL) loss, making the model fit to the annotations of each rater. The second term defines the KL-divergence between the posterior distribution $q(Z|r)$ and the prior $p(Z)$ and works as a regularization of the latent distributions. The factor $\lambda$ weights the regularization term and is set to $0.0005$ to balance the magnitudes of the log-likelihood and the KL (we will check the robustness of this hyperparameter in Section 4.5).
While the KL term can be optimized analytically, the log likelihood term must be approximated.
We use the reparametrization trick \cite{kingma_variational_2015} to split each probabilistic sample $\tilde{z}^r$ into its probabilistic component and deterministic parameters $\mu^r$ and $\Sigma^r$. These parameters can be optimized by backpropagation of gradients, together with the CNN parameters $\omega$ and $\theta$.
For numerical stability, we train the covariance matrix parameters by using the lower triangular matrix $L$ of the Cholesky decomposition  $\Sigma^r = {L^r} {L^r}^\top$ . 
The log-likelihood can be optimized  with standard methods like the categorical cross-entropy. We found that the general dice loss \cite{sudre_generalised_2017} leads to better results, so all final results are reported with this loss.

We use the Adam optimizer \cite{kingma_adam_2015} for $100$ epochs with a learning rate of $0.0001$. The model parameters $\mu^r, \Sigma^r$ are optimized with a higher learning rate of $\nu=0.02$, because else the gradient was not strong enough to properly learn the rater distributions. We tested $\nu=0.01, 0.02, 0.04$ and include the results in section \ref{sec:ablation_studies}. Both learning rates are decreased after $40$ epochs by dividing them by $1.1$ in each epoch.

\subsection{Predicting}
For a test image $x^*$, the predictive gold distribution can again be obtained by drawing Monte-Carlo samples
\begin{equation}
        \tilde{s}_{j}^*|x^*, r = f_{\theta}(v^*, \tilde{z_j}) \text{; } \tilde{z_j} \sim q(z|r=M+1)
\end{equation}
with $j=1,..,K$ indexing the MC samples and $q(z|r=M+1)$ representing the gold distribution as described in section \ref{sec:proposed_model}. The \textbf{mean} of these samples provides the segmentation hypothesis that approximates the expert agreements. The \textbf{variance} of the samples indicates uncertainties in the prediction. 

Furthermore, the model is able to simulate \textit{intra-observer variations} of rater $r'$ by drawing multiple samples of the distribution $\tilde{z}_j' \sim q(z|r=r')$ for the final prediction. The \textit{inter-observer variations} between rater $r'$ and $r''$ can be simulated by using samples $\tilde{z}_j' \sim q(z|r=r')$ and  $\tilde{z}_k'' \sim q(z|r=r'')$ and finally taking the mean of both output distributions. 

The model can therefore simulate \textbf{expert opinions} for a given test image. Other AI methods typically aggregate the expertise provided by all annotators to make predictions (e.g., using STAPLE, Prob U-Net). However, in such approaches, the knowledge of highly specialized experts can be diluted or lost among the less experienced annotators' knowledge. In our framework, we provide consistent predictions for each individual expert, thereby preserving their unique expertise and contributions.

\section{Experiments} \label{sec:experiments}
In several experiments we demonstrate that the uncertainty estimation of the model indicates areas of false predictions (\ref{sec:uncertainty}), the model is able to capture the inter and intra-observer variations (\ref{sec:variations}) and outperforms other related methods (\ref{sec:model_comparison}). Additionally, we analyze the robustness to hyperparameters (\ref{sec:ablation_studies}), required resources (\ref{sec:required_resources}), and limitations (\ref{sec:limitations}).
\subsection{Datasets} \label{sec:datasets}
% Why these datasets? Because they have rater information.
% Details: Number of images, pixels, annotations, train/test splits, preprocessing, augmentation

For empirical validation, three public histopathological datasets were used. The first dataset, “Gleason 2019" \cite{nir_automatic_2018} was published as a MICCAI grand challenge for pathology and includes 333 Tissue Micro Arrays (TMA) of prostate cancer, labeled by 6 different pathologists. The TMAs were scanned with a magnification of 40x and have a size of approximately $4000\times4000$ pixels. Of the 333 images, 244 are publicly available with labels (the test annotations of the challenge are not available). Each pathologist annotated between 61 and 241 TMAs with segmentation masks and the gold labels were obtained using the STAPLE algorithm \cite{warfield_simultaneous_2004}, following the original work of the dataset \cite{nir_automatic_2018}. We resize all images to $1024\times1024$ pixels and create 4 cross-validation splits. 

The second dataset, which we will refer to as “Arvaniti TMA" was published in 2018 \cite{arvaniti_automated_2018} and includes a total of 886 TMAs of prostate cancer of which 245 images were annotated by two pathologists (while the other images only have annotations of one pathologist and are therefore discarded in our study). The TMAs were scanned with a magnification of 40x but the scanned area is smaller than for the Gleason19 dataset. The images have a resolution of $3100\times3100$ pixels and we resize them to $512\times512$ such that the magnification matches the resized images of the Gleason 2019 dataset. Again, we split the dataset into 4 cross-validation splits for the experimental setup. 

For the classification of prostate cancer, the tissue is segmented in the Gleason Grading (GG) scheme. The classes are 'Non-cancerous' (NC), 'Gleason 3' (G3), 'Gleason 4' (G4), and 'Gleason 5' (G5) depending on the architectural growth patterns of the tumor \cite{silva-rodriguez_going_2020, silva-rodriguez_proportion_2022}. To visualize the segmentations we use the colors: green for NC, yellow for G3, orange for G4, and red for G5. For the evaluation of algorithms for prostate cancer classification, previous works used the Cohen's kappa coefficient \cite{nir_automatic_2018, arvaniti_automated_2018, silva-rodriguez_going_2020} which measures the agreement of two raters or a rater and an AI model. To compare to previously reported results for the two datasets, we use the unweighted Cohen's kappa $\kappa$ for the Gleason 2019 dataset \cite{nir_automatic_2018} and the quadratic weighted Cohen's kappa $\kappa$ for the Arvaniti TMA dataset \cite{arvaniti_automated_2018}. The main difference is that the quadratic kappa takes the class order into account and weighs the errors based on the quadratic distance of the predicted and the real class.

The third dataset contains 151 WSIs for breast cancer segmentation that were sliced into 11,836 patches of 512x512 pixels annotated by 25 raters \cite{amgad_structured_2019, lopez-perez_crowdsourcing_2023}. We will refer to this dataset as “bc segmentation". The tissue was segmented into  “tumor", “inflammation", “necrosis", “stroma", and “other". Here, the gold labels were obtained by an actual discussion of experts. We use the predefined train/validation/test splits \cite{lopez-perez_crowdsourcing_2023}.

For all datasets we use image augmentation with the albumentations library \cite{buslaev_albumentations_2020} by applying random flip, rotation, shear, zoom, blur, and shifts in brightness, contrast, hue, and saturation. This leads to a broad range of realistic transformations of the image to avoid overfitting.

\begin{figure}[t]
     \begin{center}
     \subfloat[Image\label{fig:unc_1}]{
         \centering
         \includegraphics[ width=0.13\linewidth]{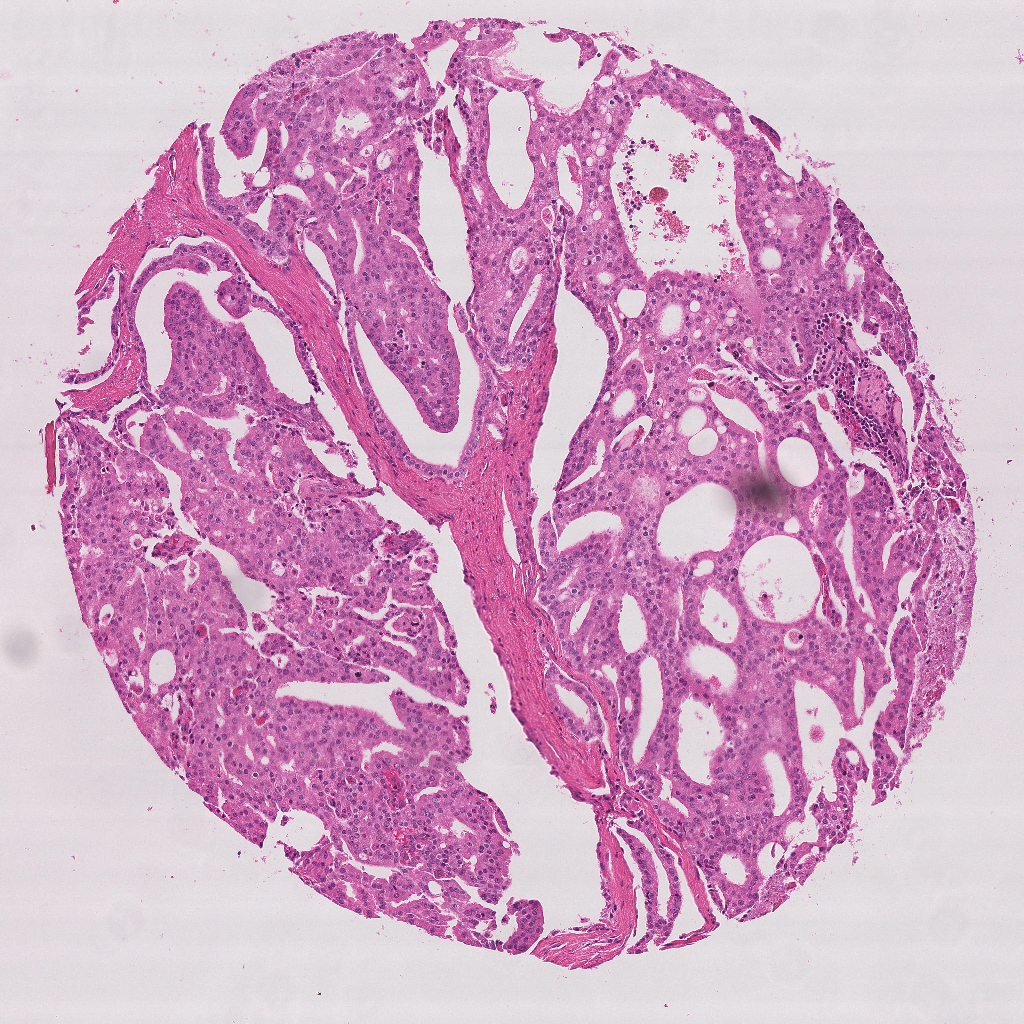}
     }
    \subfloat[GT Gold\label{fig:unc_2}]{
         \centering
         \includegraphics[ width=0.13\linewidth]{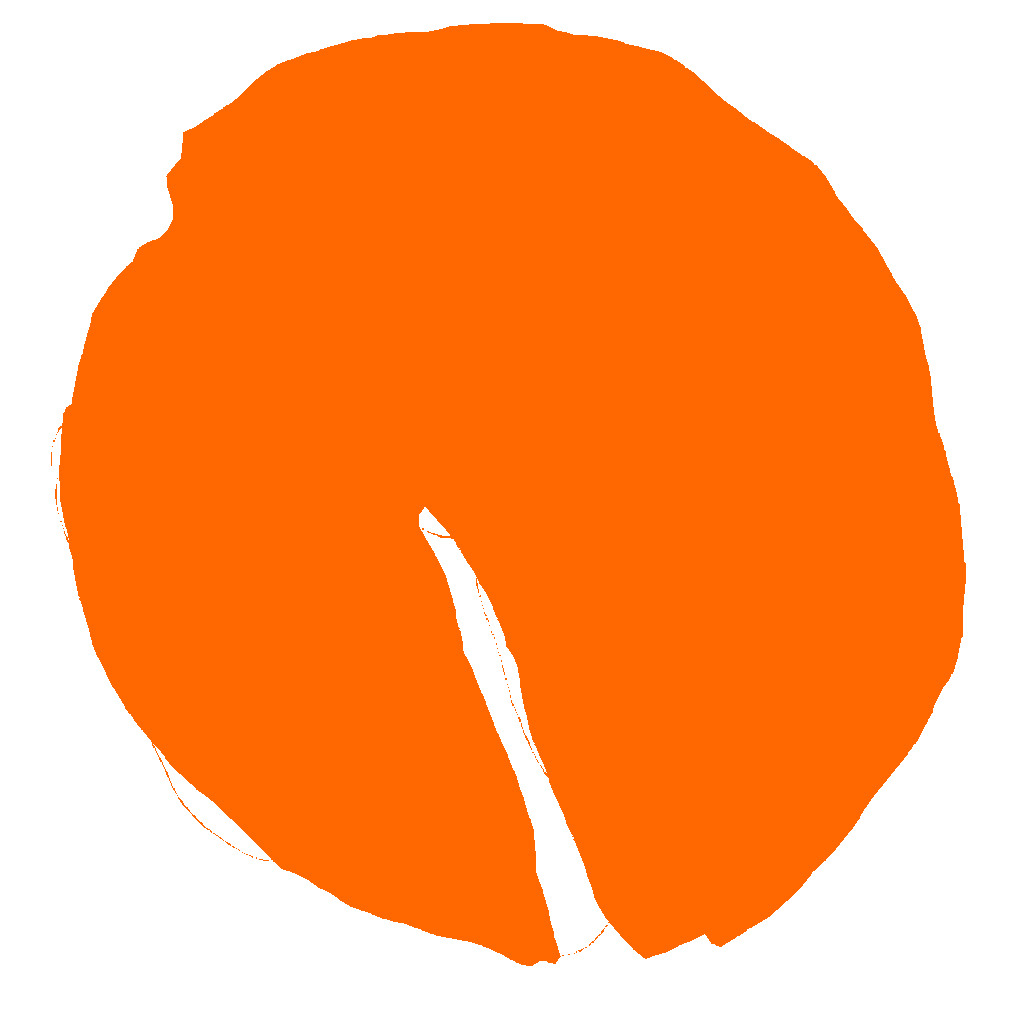}
     }
    \subfloat[Pred. Gold\label{fig:unc_3}]{
         \centering
         \includegraphics[ width=0.13\linewidth]{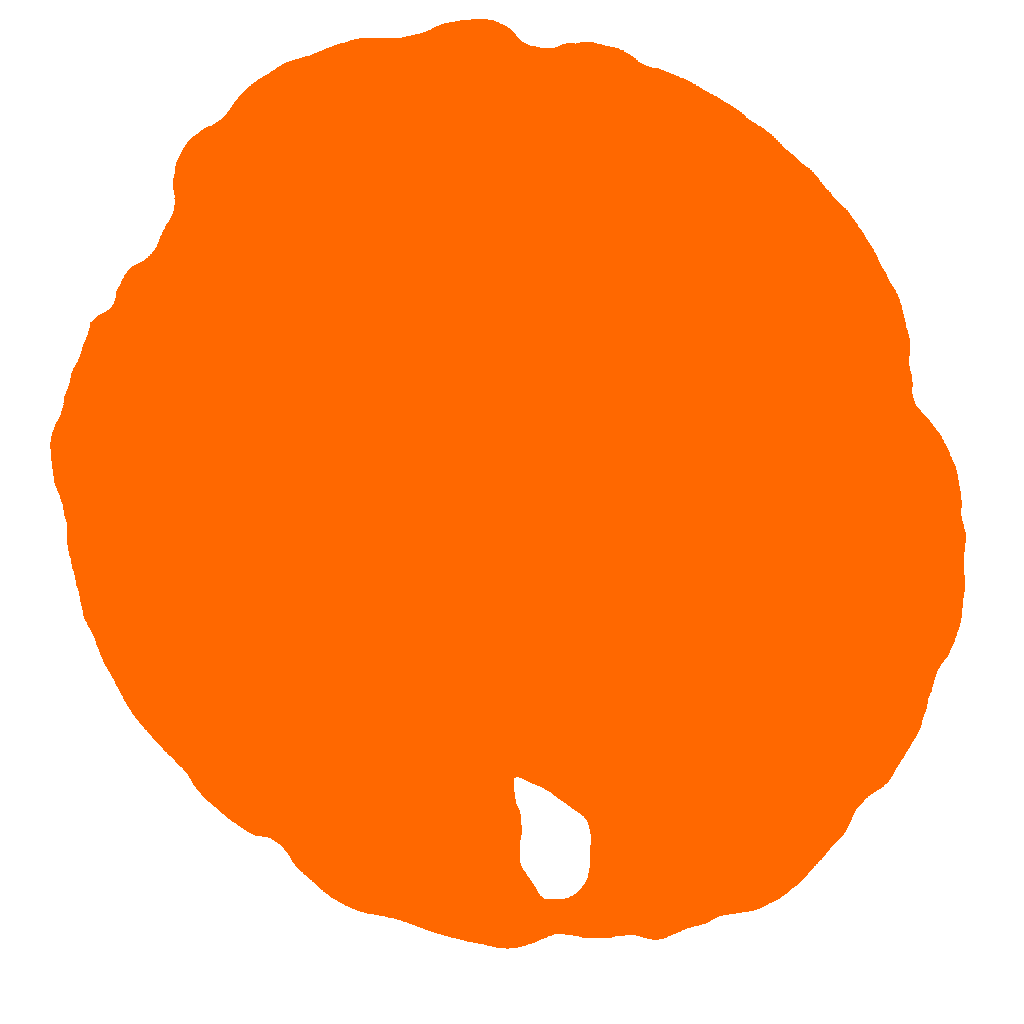}
     }
    \subfloat[Uncert.\label{fig:unc_4}]{
         \centering
         \includegraphics[ width=0.13\linewidth]{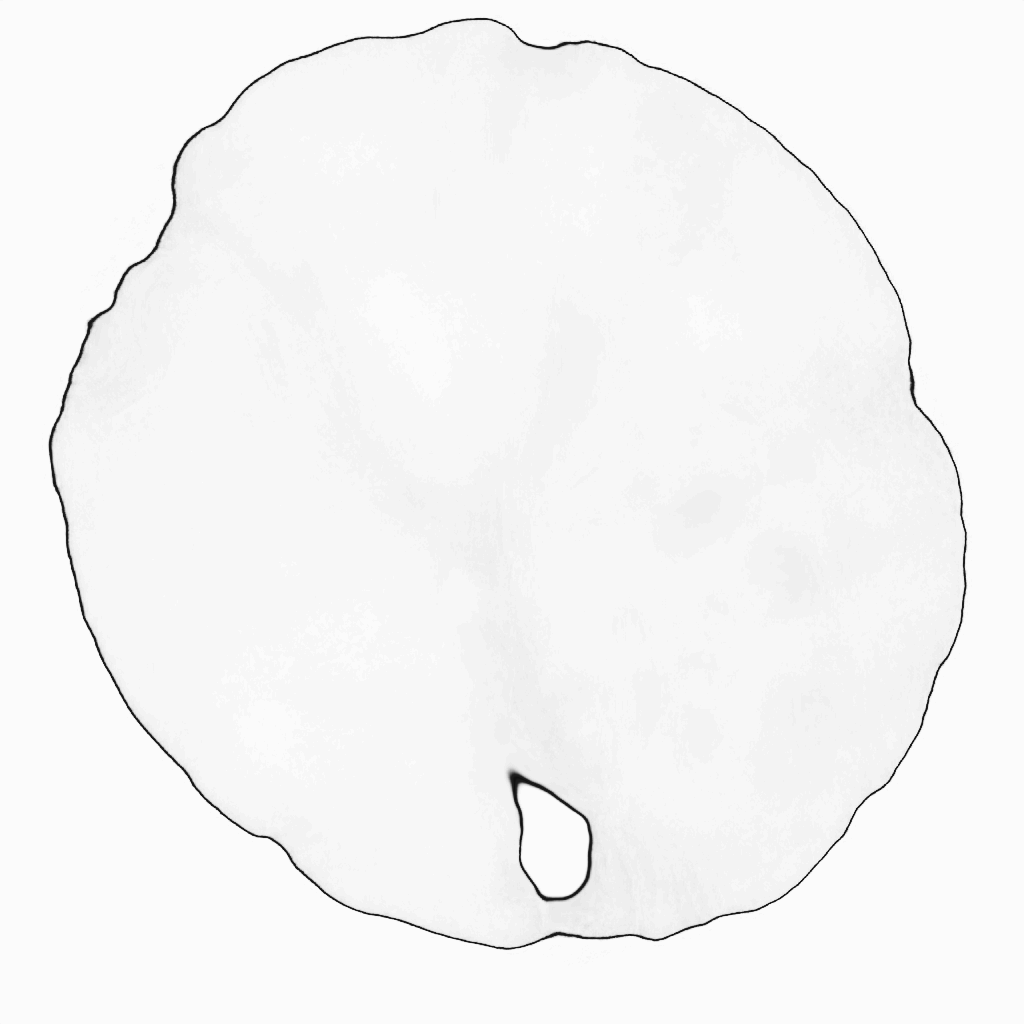}
     }\\
     \subfloat[Image\label{fig:unc_5}]{
         \centering
         \includegraphics[ width=0.13\linewidth]{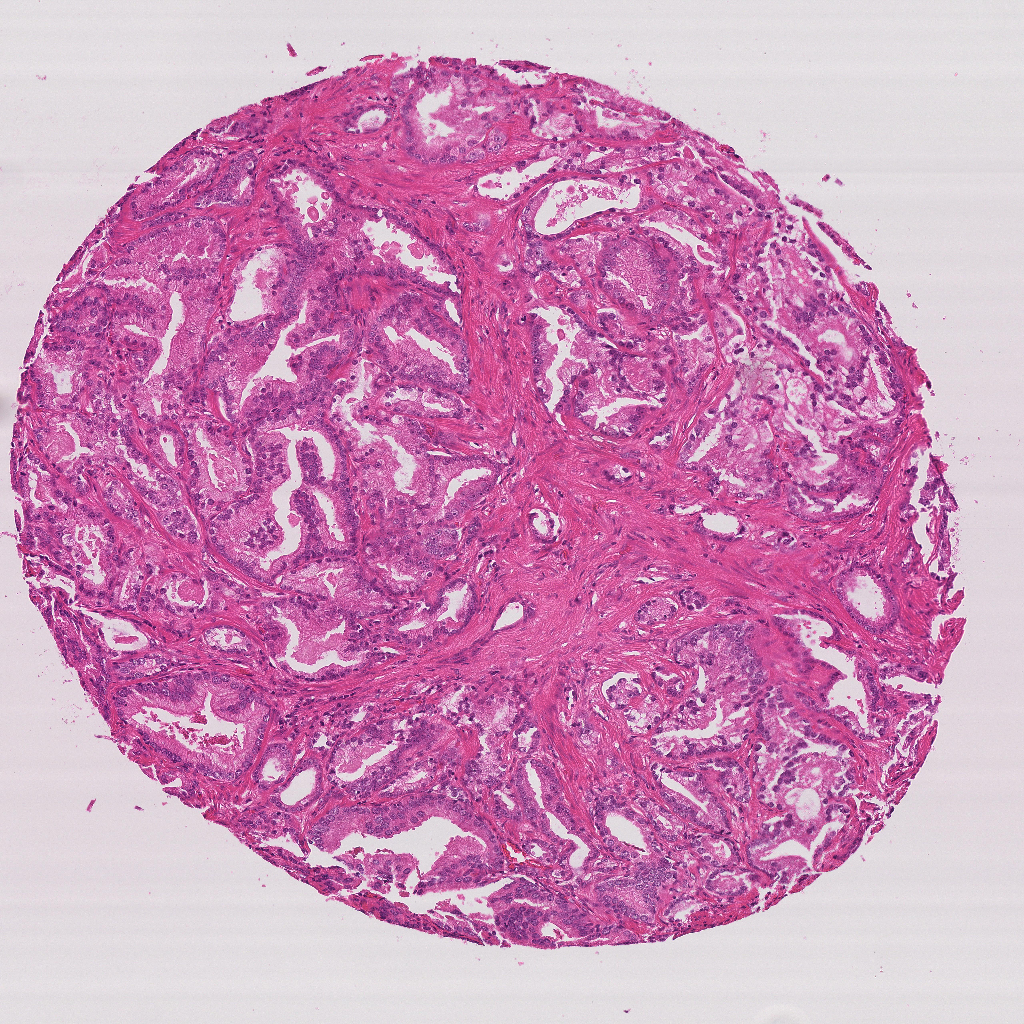}
     }
    \subfloat[GT Gold\label{fig:unc_6}]{
         \centering
         \includegraphics[ width=0.13\linewidth]{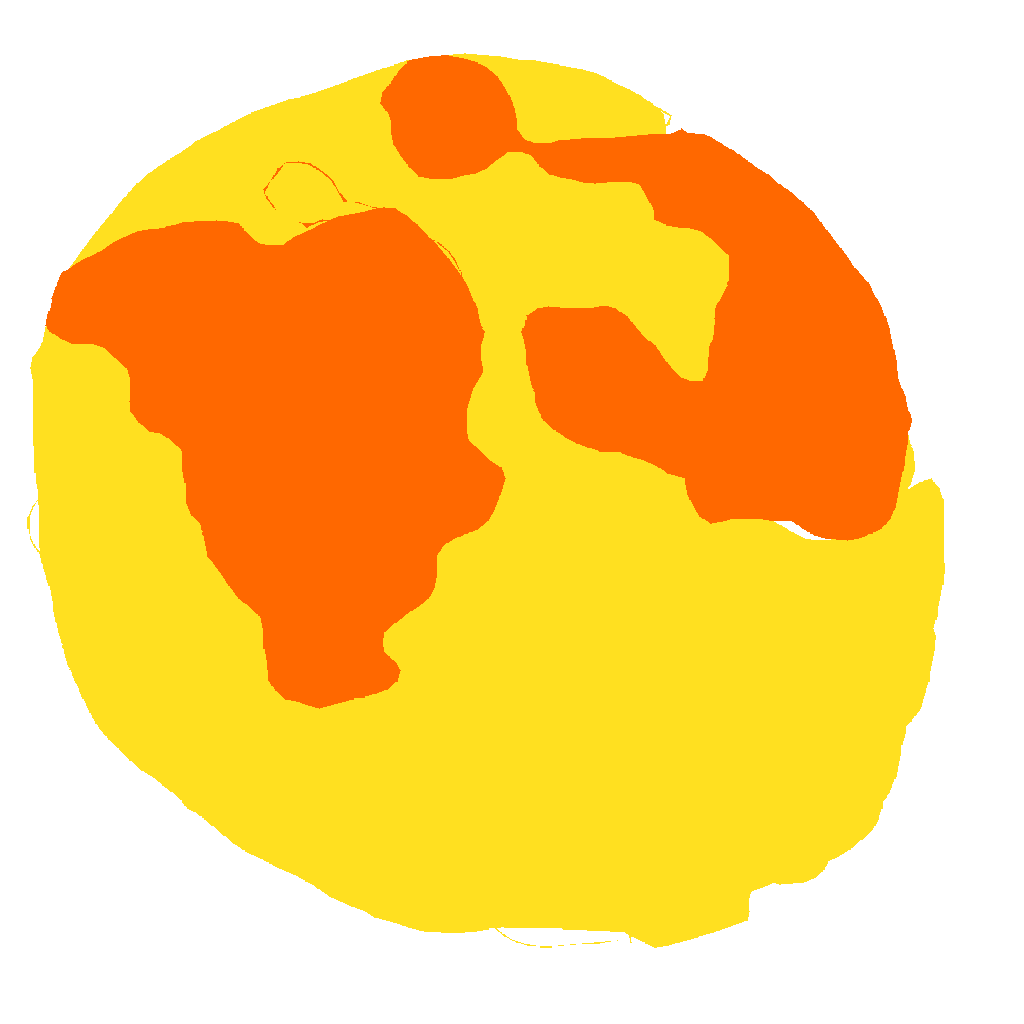}
     }
    \subfloat[Pred. Gold\label{fig:unc_7}]{
         \centering
         \includegraphics[ width=0.13\linewidth]{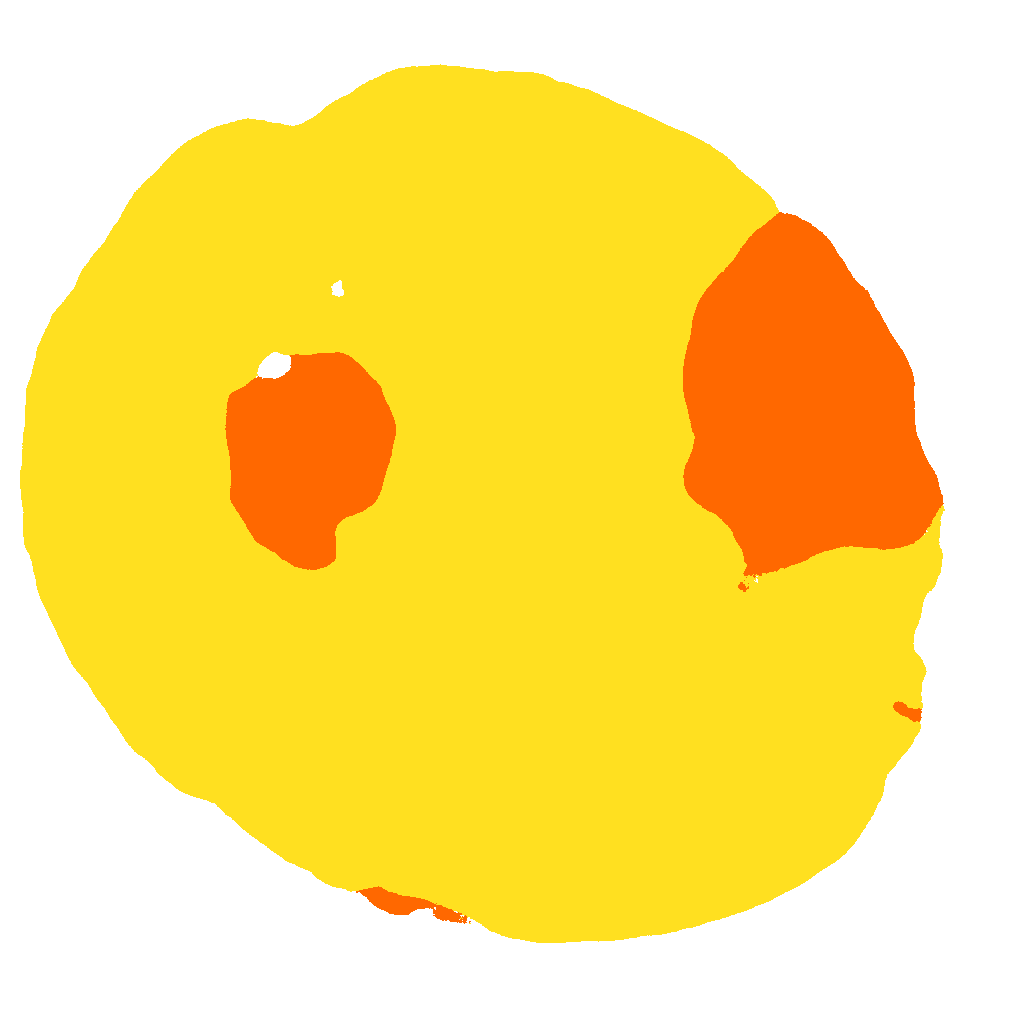}
     }
    \subfloat[Uncert.\label{fig:unc_8}]{
         \centering
         \includegraphics[ width=0.13\linewidth]{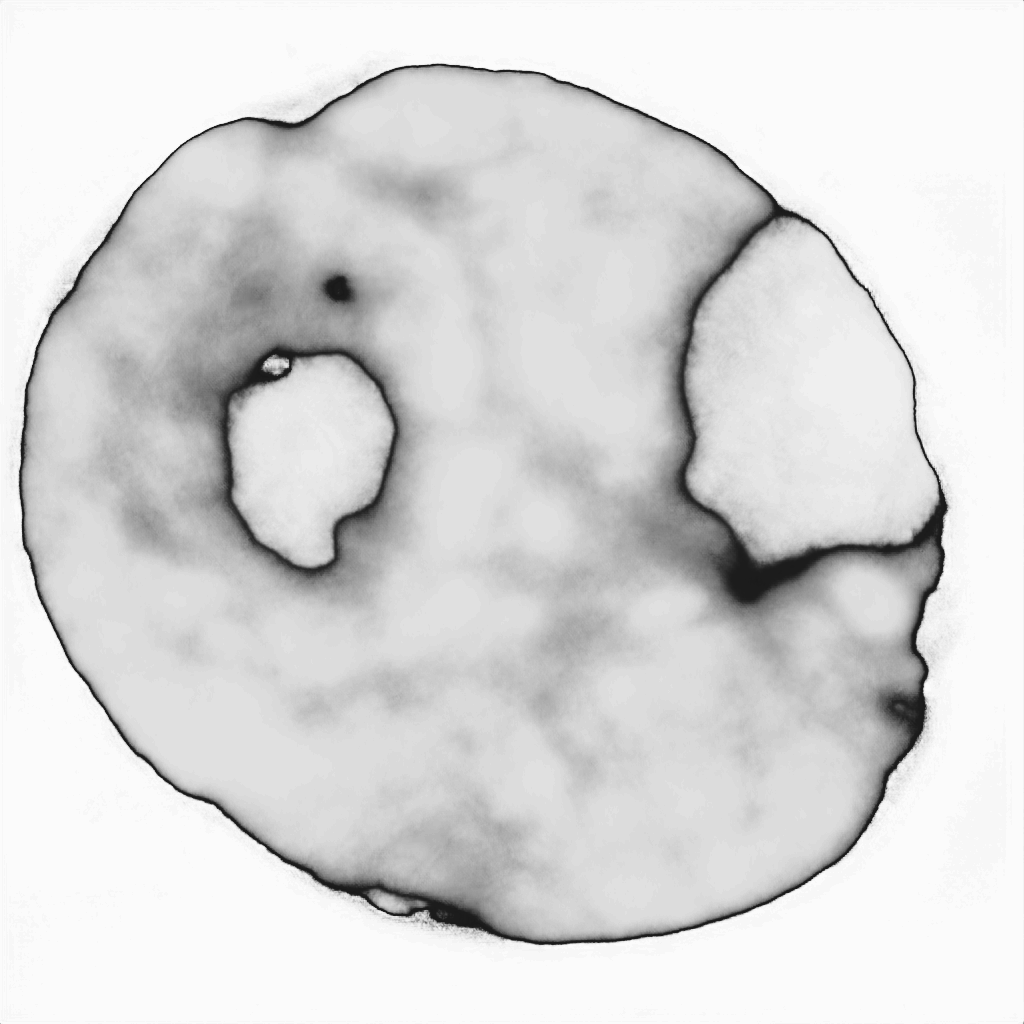}
     }
     \end{center}
    \caption{Gold prediction and uncertainty of the Pionono model. The first row shows a confident prediction as the uncertainty in \ref{fig:unc_4} is low (white) for almost all the area. Indeed, the segmentation prediction \ref{fig:unc_3} is very accurate, see the ground truth (GT) \ref{fig:unc_2}. The second row shows an example of an uncertain prediction. Some parts of the area classified as G3 (yellow) in \ref{fig:unc_7} are labeled as G4 in the ground truth \ref{fig:unc_6}. These areas are estimated with a high uncertainty (black) in \ref{fig:unc_8}, warning that these predictions are unreliable. }
        \label{fig:uncertainty_estimation}
\end{figure}
\subsection{Uncertainty estimation} \label{sec:uncertainty}
% The model uncertainty estimation is working well and the output is helpful
The proposed model provides probabilistic predictions that allow an accurate assessment of the predictive uncertainty. Fig. \ref{fig:uncertainty_estimation} shows the model predictions and uncertainties obtained with the gold distribution as described in section \ref{sec:proposed_model}. For the first image (\ref{fig:unc_1}), the prediction of the model (\ref{fig:unc_3}) is accurate and matches the real gold annotation (\ref{fig:unc_2}) very well. The uncertainty (\ref{fig:unc_4}) for this predictions is low (white), which means that there is a low risk of a wrong prediction. Therefore, the model correctly indicates that this prediction is reliable. For the second image \ref{fig:unc_5}, some areas that are predicted as G3 (yellow) in (\ref{fig:unc_7}) are actually G4 in the ground truth gold prediction (\ref{fig:unc_6}). The model's uncertainty estimation indicates that this prediction is not reliable: the misclassified areas are marked with a high uncertainty (dark) in the image (\ref{fig:unc_8}). Therefore, the probabilistic output adds valuable information to the diagnostic process. It estimates if a prediction is reliable - or unreliable and should be double-checked.

\subsection{Inter- and Intra-observer Variation} \label{sec:variations}

% The model captures variations and can simulate them during prediction.
The Pionono model is able to capture the inter- and intra-observer variability. This accurate probabilistic modeling of the annotations does not only improve the predictive results (see section \ref{sec:model_comparison}), but also allows to simulate specific experts at test time.
% We want to make sure with the experiments, that the raters are modeled correctly
In this section, we empirically show that the model learns the different label behaviors of the raters and is able to reproduce them. 

\begin{figure}[t]
     \begin{center}
    \subfloat[\label{fig:agreement}]{

    \includegraphics[ width=0.35\linewidth]{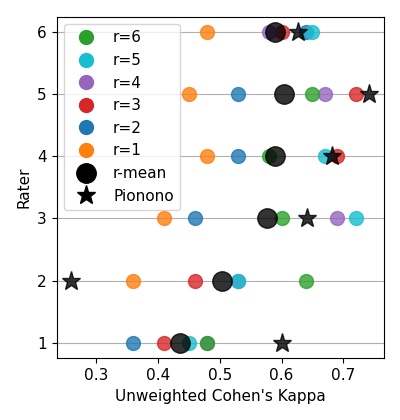}
    }
    \subfloat[\label{fig:distribution}]{
    \includegraphics[ width=0.35\linewidth]{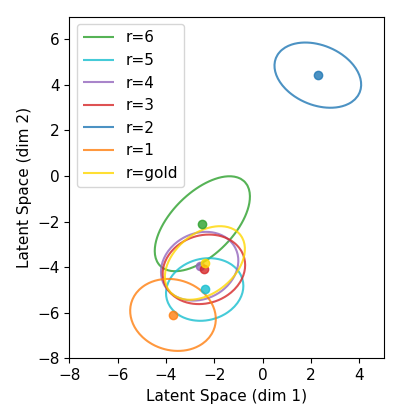}
    }
    \end{center}
    \caption{Analyzing the labeling behaviour. In Fig. (a) the agreement of each rater with all other raters is depicted, measured by the unweighted Cohen's Kappa of the true labels \cite{nir_automatic_2018}. The mean agreement of each rater with all other raters is represented by a black dot and the agreement with Piononos test predictions, simulating the corresponding rater, by a star. This confirms that the model accurately models each rater, reaching even a higher agreement than the other raters average, except for rater 2. Fig. (b) shows the first two dimensions of the posterior distributions $q(z|r)$ with mean and covariance after training. The distributions of raters $3,4,5$ and $6$ overlap significantly with the gold distribution and with each other, indicating a similar labeling behaviour. Indeed, these raters show the highest labeling agreement of true labels, as observed in (a).}
    \label{fig:inter_obsever_agreement}
\end{figure}

% We plot the agreements between the raters, the mean agreement of the other raters and pionono.
In Fig. \ref{fig:agreement} we plot the \textit{inter-observer variations} between the raters. The figure shows that there is indeed a high variability among the raters, with a Cohen's kappa ranging from $0.36$ to $0.72$.  The simulated test predictions by Pionono show a higher agreement with each rater than the average agreement of the other raters, except for rater $2$. For two raters ($1$ and $5$), the simulated predictions of Pionono are even more than $15$ percentage points higher than the average rater agreement. We also measured the IoU metric, which was $0.574, 0.540, 0.619, 0.649, 0.692, 0.507$, for the 6 raters respectively, compared to a mean inter-pathologist IoU of $0.361$.
The results confirm that most raters are modeled with high accuracy. 

\begin{figure}[H]
\captionsetup[subfigure]{font=scriptsize,labelfont=scriptsize}
     \begin{center}
     \hspace{-0.116\linewidth}
     \subfloat[Image \label{fig:inter_a}]{
         \centering
         \includegraphics[ width=0.08\linewidth]{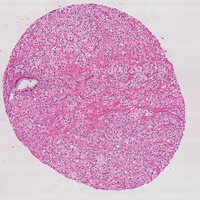}
     }
     \hspace{0.12\linewidth}
    \vrule
    \hspace{0.01\linewidth}
    \subfloat[Image \label{fig:inter_b}]{
         \centering
         \includegraphics[ width=0.08\linewidth]{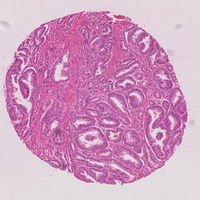}
     }
     \hspace{0.2\linewidth}
     \\
    \subfloat[GT $r=1$\label{fig:inter_c}]{
         \centering
         \includegraphics[ width=0.08\linewidth]{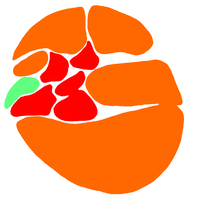}
     }
     \hspace{0.01\linewidth}
     \subfloat[Pred $r=1$\label{fig:inter_d}]{
         \centering
         \includegraphics[ width=0.08\linewidth]{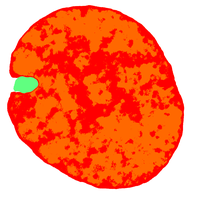}
    }  
    \hspace{0.01\linewidth}
    \vrule
    \hspace{0.01\linewidth}
    \subfloat[GT $r=1$\label{fig:inter_e}]{
         \centering
         \includegraphics[ width=0.08\linewidth]{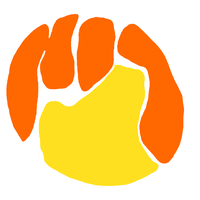}
     }
    \hspace{0.01\linewidth}
    \subfloat[Pred $r=1$\label{fig:inter_f}]{
         \centering
         \includegraphics[ width=0.08\linewidth]{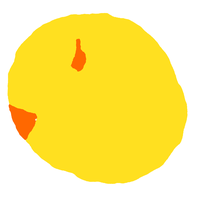}
    }   
    \\
    \subfloat[GT $r=2$\label{fig:inter_g}]{
         \centering
         \includegraphics[ width=0.08\linewidth]{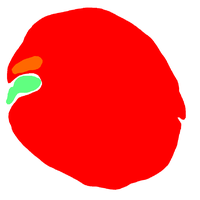}
     }
     \hspace{0.01\linewidth}
    \subfloat[Pred $r=2$\label{fig:inter_h}]{
         \centering
         \includegraphics[ width=0.08\linewidth]{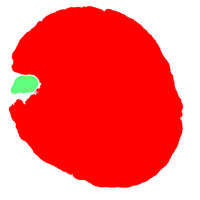}
    } 
    \hspace{0.01\linewidth}
    \vrule
    \hspace{0.01\linewidth}
    \subfloat[GT $r=2$\label{fig:inter_i}]{
         \centering
         \includegraphics[ width=0.08\linewidth]{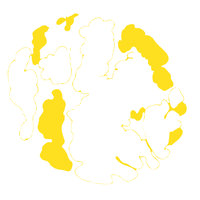}
     }
     \hspace{0.01\linewidth}
    \subfloat[Pred $r=2$\label{fig:inter_j}]{
         \centering
         \includegraphics[ width=0.08\linewidth]{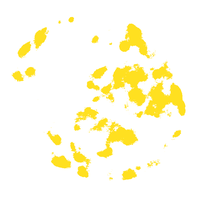}
    }      
    \\
     \subfloat[GT $r=3$\label{fig:inter_k}]{
         \centering
         \includegraphics[ width=0.08\linewidth]{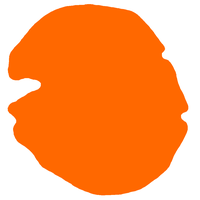}
     }
     \hspace{0.01\linewidth}
     \subfloat[Pred $r=3$\label{fig:inter_l}]{
         \centering
         \includegraphics[ width=0.08\linewidth]{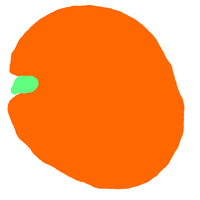}
    } 
    \hspace{0.01\linewidth}
    \vrule
    \hspace{0.01\linewidth}
     \subfloat[GT $r=3$\label{fig:inter_m}]{
         \centering
         \includegraphics[ width=0.08\linewidth]{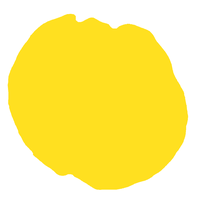}
     }  
     \hspace{0.01\linewidth}
     \subfloat[Pred $r=3$\label{fig:inter_n}]{
         \centering
         \includegraphics[ width=0.08\linewidth]{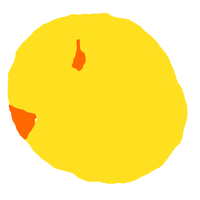}
    } 
    \\
    \subfloat[GT $r=4$\label{fig:inter_o}]{
         \centering
         \includegraphics[ width=0.08\linewidth]{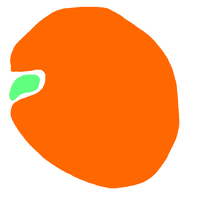}
     }
     \hspace{0.01\linewidth}
    \subfloat[Pred $r=4$\label{fig:inter_p}]{
         \centering
         \includegraphics[ width=0.08\linewidth]{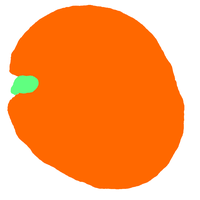}
    } 
    \hspace{0.01\linewidth}
    \vrule
    \hspace{0.01\linewidth}
    \subfloat[GT $r=4$\label{fig:inter_q}]{
         \centering
         \includegraphics[ width=0.08\linewidth]{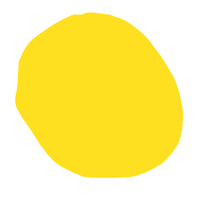}
     }
     \hspace{0.01\linewidth}
    \subfloat[Pred $r=4$\label{fig:inter_r}]{
         \centering
         \includegraphics[ width=0.08\linewidth]{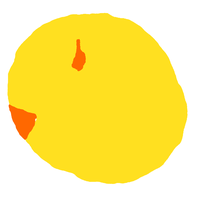}
    } 
    \\
    \subfloat[GT $r=5$\label{fig:inter_s}]{
         \centering
         \includegraphics[ width=0.08\linewidth]{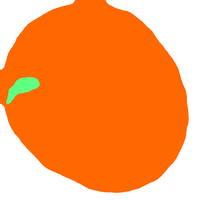}
     }
     \hspace{0.01\linewidth}
     \subfloat[Pred $r=5$\label{fig:inter_t}]{
         \centering
         \includegraphics[ width=0.08\linewidth]{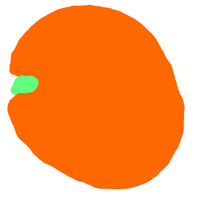}
    } 
    \hspace{0.01\linewidth}
    \vrule
    \hspace{0.01\linewidth}
    \subfloat[GT $r=5$\label{fig:inter_u}]{
         \centering
         \includegraphics[ width=0.08\linewidth]{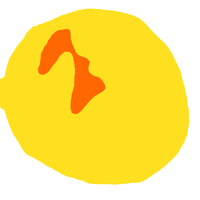}
     }
     \hspace{0.01\linewidth}
    \subfloat[Pred $r=5$\label{fig:inter_v}]{
         \centering
         \includegraphics[ width=0.08\linewidth]{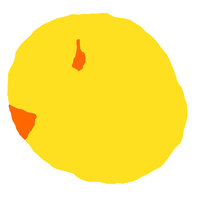}
    } 
    \\
    \subfloat[GT $r=6$\label{fig:inter_w}]{
         \centering
         \includegraphics[ width=0.08\linewidth]{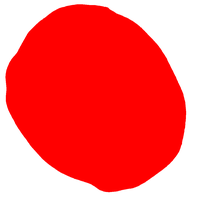}
     }   
     \hspace{0.01\linewidth}
    \subfloat[Pred $r=6$\label{fig:inter_x}]{
         \centering
         \includegraphics[ width=0.08\linewidth]{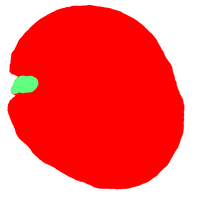}
    } 
    \hspace{0.01\linewidth}
    \vrule
    \hspace{0.01\linewidth}
    \subfloat[GT $r=6$\label{fig:inter_y}]{
         \centering
         \includegraphics[ width=0.08\linewidth]{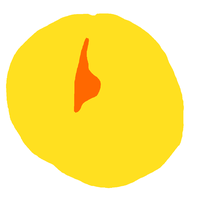}
     }   
     \hspace{0.01\linewidth}
    \subfloat[Pred $r=6$\label{fig:inter_z}]{
         \centering
         \includegraphics[ width=0.08\linewidth]{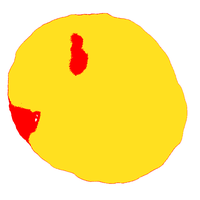}
    } 
    \end{center}
    \caption{
   Inter-observer variations estimated by Pionono. For two test images we depict the ground truth (GT) segmentations of all raters and the predicted segmentations, simulating each rater. The proposed model is able to simulate certain labeling behaviour like the tendency of assigning class G5 (red) for raters 2 and 6 (see g and w) where other raters assigned G4 (orange). Furthermore, the model captures the under-segmention by rater 2 (see i).
    }
    \label{fig:inter_observer_var}
\end{figure}

Fig. \ref{fig:distribution} shows the posterior distributions $q(z|r)$ of the proposed model, encoding the labeling behavior of each rater. The following observations confirm, that these learned distributions approximate well the real-world labeling behavior of the raters: (i) The four raters $3,4,5$ and $6$ show a high overlap of the distributions and corresponding to a high labeling agreement shown in Fig. \ref{fig:agreement}. (ii) The gold distribution (simulating raters agreement) overlaps significantly with the distribution of these four raters. (iii) The distribution of rater $2$ is far away from all other distributions. This rater shows a different labeling behavior due to frequent under-segmentation of images, assigning the 'background' class to areas that contain tissue. (iv) Raters $1$ and $6$ often deviate from the other raters, especially for the differentiation of classes G3 and G4. Their distribution accordingly has a smaller overlap with the gold distribution and the other raters. Fig. \ref{fig:inter_observer_var} shows some visual image examples of Pionono test predictions, simulating each rater $r$ by drawing samples from the corresponding distribution $q(z|r)$ and then taking the mean of the output samples. The examples confirm that the rater differences are modeled well.

\begin{figure}[t]
\captionsetup[subfigure]{font=scriptsize,labelfont=scriptsize}
     \begin{center}
     \subfloat[Image\label{fig:intra_0}]{
         \centering
         \includegraphics[ width=0.08\linewidth]{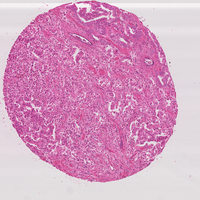}
     }
     \hspace{0.01\linewidth}
     \subfloat[Pred $r=6$\label{fig:intra_2}]{
         \centering
         \includegraphics[ width=0.08\linewidth]{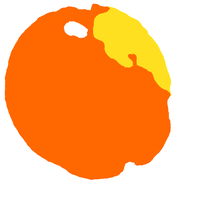}
     }
     \hspace{0.01\linewidth}
    \vrule
    \hspace{0.01\linewidth}
     \subfloat[Image\label{fig:intra_7}]{
         \centering
         \includegraphics[ width=0.08\linewidth]{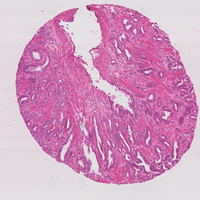}
     }
     \hspace{0.01\linewidth}
    \subfloat[Pred $r=2$\label{fig:intra_9}]{
         \centering
         \includegraphics[ width=0.08\linewidth]{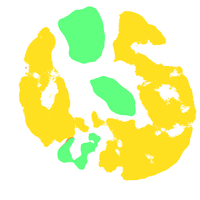}
     }
    \\
    \subfloat[GT $r=6$\label{fig:intra_1}]{
         \centering
         \includegraphics[ width=0.08\linewidth]{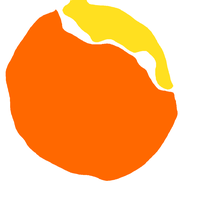}
     }
     \hspace{0.01\linewidth}
     \subfloat[Pred $r=6$\label{fig:intra_3}]{
         \centering
         \includegraphics[ width=0.08\linewidth]{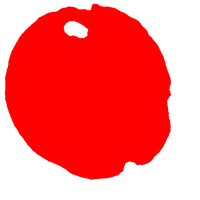}
     }
     \hspace{0.01\linewidth}
    \vrule
     \hspace{0.01\linewidth}
    \subfloat[GT $r=2$\label{fig:intra_8}]{
         \centering
         \includegraphics[ width=0.08\linewidth]{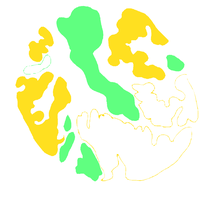}
     }
     \hspace{0.01\linewidth}
     \subfloat[Pred $r=2$\label{fig:intra_10}]{
         \centering
         \includegraphics[ width=0.08\linewidth]{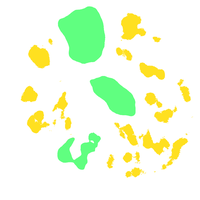}
     }
     \\
     
     \hspace{0.105\linewidth}
      \subfloat[Pred $r=6$\label{fig:intra_4}]{
         \centering
         \includegraphics[ width=0.08\linewidth]{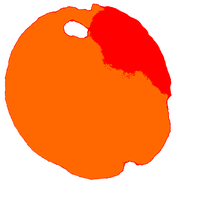}
     }
     \hspace{0.01\linewidth}
     \vrule
     \hspace{0.12\linewidth}
    \subfloat[Pred $r=2$\label{fig:intra_11}]{
         \centering
         \includegraphics[ width=0.08\linewidth]{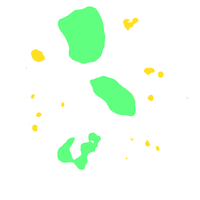}
     }
     \\
     
    \hspace{0.105\linewidth}
    \subfloat[Pred $r=6$\label{fig:intra_6}]{
         \centering
         \includegraphics[ width=0.08\linewidth]{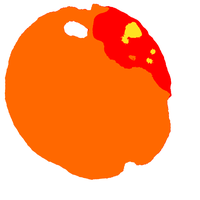}
     }
     \hspace{0.01\linewidth}
     \vrule
     \hspace{0.12\linewidth}
    \subfloat[Pred $r=2$\label{fig:intra_12}]{
         \centering
         \includegraphics[ width=0.08\linewidth]{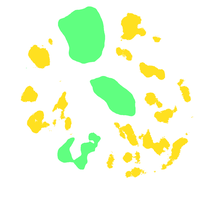}
     }
    \\
    
    \hspace{0.105\linewidth}
    \subfloat[Pred $r=6$\label{fig:intra_5}]{
         \centering
         \includegraphics[ width=0.08\linewidth]{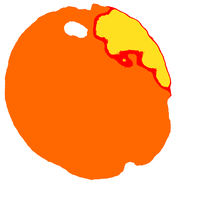}
     }
    \hspace{0.01\linewidth}
    \vrule
    \hspace{0.12\linewidth}
    \subfloat[Pred $r=2$\label{fig:intra_12}]{
         \centering
         \includegraphics[ width=0.08\linewidth]{images/intra_var/pred_slide006_core105_s_5_a2.png}
     }
     \\
    \end{center}
    \caption{
    Intra-observer variations estimated by the Pionono model. For two example images we depict the true annotation of two different raters ($r=2$ and $r=6$). On the right side we show different coherent segmentation hypothesis for each rater estimated by our model. The differences in each column reflect possible intra-observer variations. The first example (a) shows that the segmentation of rater 6 might show some variations in the assigned classes. While the segmentation prediction (b) is composed by classes G3 (yellow) and G4 (orange), the segmentation sample (f) only consists of G5 (red). Indeed, this rater often assigns class G5 (red) in areas where other raters assign G4 (see Fig. \ref{fig:inter_observer_var}) such that this is a plausible hypothesis. In the second example (c) we see variations due to the under-segmentation of rater 2 in some images. Our model captures this behaviour and provides different hypothesis of more (d) or less (j) segmentation of class G3 (yellow).
    }
    \label{fig:intra_observer_var}
\end{figure}

Next, we analyze the \textit{intra-observer variations}. As the dataset does not contain multiple annotations of the same rater for the same image, the assessment of this quality is more difficult. Still, certain intra-observer variability can be assessed by observing the general labeling behavior of one annotator. For example, rater $6$ tends to over-assign class G5 (red), and rater $2$ tends to not segment all image parts that contain tissue. Interestingly, these intra-observer variations are present in the model predictions when multiple samples are drawn from their corresponding distribution. Fig. \ref{fig:intra_observer_var} shows visual examples of the simulated variations of raters $2$ and $6$.

\subsection{Model Comparison} \label{sec:model_comparison}
The proposed model is compared to previously reported results and several state-of-the-art approaches (see section \ref{sec:related_work}) for medical image segmentation with labels from multiple raters. For fair comparison, we use the same backbone architecture\footnote{Only the model \textit{CM pixel} uses ResNet18 to fit on the GPU.}, epochs, learning rate, and optimizer for all experiments. We have tuned the model-specific hyperparameters to obtain the best possible results for each method. 

First, we perform experiments with the Gleason 2019 dataset with a 4-fold crossvalidation. For comparison with previous works, we report the unweighted Cohen's kappa metric comparing gold predictions with gold ground truth. Additionally, we report the accuracy. As the results in Table \ref{tab:gleason_19} show, the proposed Pionono model outperforms the previously reported results \cite{nir_automatic_2018, qiu_automatic_2022}, including the winner of the Gleason 2019 challenge \cite{qiu_automatic_2022}, by a large margin of over $20$ percentage points. This accounts for the exact modeling of the raters by Pionono, but also for the different choices of backbone architecture and other training details. Compared to other state-of-the-art methods with the same architecture and training details, Pionono still shows a considerably better performance. 

Next, we compare the generalization capabilities of the models by using the Arvaniti TMA dataset as an external test set, as reported in Table \ref{tab:external_test}. This means, that the models are trained with all images from Gleason 2019 and tested with all images from Arvaniti TMA. As the Arvaniti TMA dataset does not contain gold labels, the model's gold predictions are compared to both raters independently, as previously done by Arvaniti \etal \cite{arvaniti_automated_2018}. We observe that the model generalizes better than all other methods, achieving a higher agreement in terms of Cohen's quadratic kappa with both raters. In terms of accuracy, only 'CM global' outperforms Pionono by a small margin.

In the third experiment, we use the Arvaniti TMA dataset for training and testing with the two rater annotations. Again, the proposed model is able to outperform previously reported results as well as other state-of-the-art methods in terms of quadratic Cohen's kappa. In terms of accuracy, only the “Prob U-Net" model obtains a better result for rater 1, while Pionono reaches the best accuracy for rater 2. 

To validate the model on a different kind of data, we performed the fourth experiment on the “bc segmentation" dataset \cite{amgad_structured_2019}. The results are reported in Table \ref{tab:breast_cancer} and confirm the strong performance of the proposed Pionono model.
The results support our hypothesis that explicitly modeling the inter- and intra-observer variations improves the model's performance. Pionono takes the different labeling behavior into account during training which leads to accurate predictions.

\begin{table}
\small
\begin{center}
\begin{tabular}{rll}
\toprule
 Method      &          Unweighted $\kappa$ & Accuracy \\
\midrule
Nir \etal \cite{nir_automatic_2018}  &     $0.51$ & N.A.   \\
Qiu \etal \cite{qiu_automatic_2022} &     $0.524$ & N.A.   \\
\midrule
%superv. noisy &  0.751  & 0.833  \\
STAPLE &  $0.75 \pm 0.006$  & $0.834 \pm 0.005$  \\
Prob U-Net  &   $0.741 \pm 0.002$  & $0.83 \pm 0.001$  \\
CM global & $0.721 \pm 0.018$ & $0.814 \pm 0.012$ \\
CM pixel & $0.692 \pm 0.019$ & $0.791 \pm 0.012$ \\
Pionono  &     $\mathbf{0.758}\pm 0.011$   & $\mathbf{0.84}\pm 0.007$   \\
\bottomrule
\end{tabular}
\end{center}

\caption{Cohens Kappa Comparison for the 4-fold crossvalidation experiment of the Gleason 2019 dataset, reported by mean and standard error.} \label{tab:gleason_19}
\end{table}

\begin{table}
\small
\begin{center}
\begin{tabular}{rll}
\toprule
\textit{Rater 1} \ \
 Method      &          Quadratic $\kappa$ & Accuracy\\
\midrule
STAPLE &  $0.629 \pm 0.002$ & $0.718    \pm 0.001$\\
Prob U-Net  &   $0.629 \pm 0.005$& $0.73 \pm 0.003$\\
CM global & $0.624 \pm 0.003$ & $0.728 \pm 0.002$\\
CM pixel & $0.618\pm 0.007$  & $0.72\pm 0.004 $  \\
Pionono  &     $\mathbf{0.641}\pm 0.006$   & $\mathbf{0.736}\pm 0.004$    \\
\midrule
\textit{Rater 2} \ \
 Method      &          Quadratic $\kappa$ & Accuracy\\
\midrule
STAPLE &  $0.563\pm 0.002$  & $0.621\pm 0.002$  \\
Prob U-Net  &   $0.56\pm 0.003$ & $0.626 \pm 0.006$  \\
CM global  & $0.557\pm 0.003$ &  $\mathbf{0.638}\pm 0.006$  \\
CM pixel  & $0.551\pm 0.006$ & $0.626\pm 0.008$\\
Pionono  &     $\mathbf{0.569}\pm 0.005$  & $0.633\pm 0.005$   \\
\bottomrule
\end{tabular}
\end{center}

\caption{Cohens Kappa Comparison with the two raters of the Arvaniti TMA dataset trained on the Gleason 2019 dataset, reported by mean and standard error.} \label{tab:external_test}
\end{table}

\begin{table}
\small
\begin{center}
\begin{tabular}{rll}
\toprule
\textit{Rater 1} \ \
 Method      &          Quadratic $\kappa$ & Accuracy\\
\midrule
Arvaniti \etal \cite{arvaniti_automated_2018}  &     $0.55$ & N.A.   \\
Silva-R. \etal \cite{silva-rodriguez_going_2020}  &     $0.536$ & N.A.   \\
\midrule
supervised &  $0.658\pm 0.025$ & $0.734\pm 0.008$ \\
Prob U-Net  &   $0.697\pm 0.008$ & $\mathbf{0.762} \pm 0.004$\\
CM global  & $0.677 \pm 0.028$& $ 0.745 \pm 0.011$ \\
CM pixel  & $0.647 \pm 0.016$& $0.731\pm 0.012$\\
Pionono  &     $\mathbf{0.716}\pm 0.011$  & $0.751 \pm 0.02$  \\
\midrule
\textit{Rater 2} \ \
 Method      &          Quadratic $\kappa$ & Accuracy\\
\midrule
Arvaniti  &     $0.49$ & N.A.   \\
\midrule
supervised &  $0.521 \pm 0.014$ & $0.678 \pm 0.014$\\
Prob U-Net  &   $0.534 \pm 0.002$& $0.68 \pm 0.005$ \\
CM global & $0.533 \pm 0.022$& $0.676 \pm 0.011$\\
CM pixel & $0.508 \pm 0.013$& $0.663 \pm 0.008$\\
Pionono  &     $\mathbf{0.548} \pm 0.008$ & $\mathbf{0.697} \pm 0.012$ \\
\bottomrule
\end{tabular}
\end{center}

\caption{Cohens Kappa Comparison with the two raters of the Arvaniti TMA dataset trained and validated by 4-fold crossvalidation of the Arvaniti data, reported by mean and standard error.}
\end{table}

\begin{table}
\begin{center}
\small
\begin{tabular}{rll}
\toprule
 Method      &        Unweighted $\kappa$ & Accuracy \\
\midrule
STAPLE &  $0.647 \pm 0.003$  & $0.755 \pm 0.002$  \\
Prob U-Net  &   $0.685 \pm 0.023$  & $0.734 \pm 0.004$  \\
CM global & $0.654 \pm 0.005$ & $0.761 \pm 0.004$ \\
CM pixel & $0.689 \pm 0.010$ & $0.784 \pm 0.007$ \\
Pionono  &     $\mathbf{0.711}\pm 0.002$   & $\mathbf{0.799}\pm 0.001$   \\
\bottomrule
\end{tabular}
\end{center} 
\caption{\centering Results for the breast cancer segmentation of WSIs, reported as mean and standard error of 4 runs.} \label{tab:breast_cancer}
%\vspace{-6.5mm}
\end{table}

\subsection{Robustness to Hyperparameter Settings} \label{sec:ablation_studies}
To measure the sensitivity of the model regarding different hyperparameters, we performed studies on the 4-fold cross-validation experiment of the Gleason 19 dataset. Table \ref{tab:ablation_studies} shows that the model is robust to variations of all analyzed hyperparameters. 
We observe minor performance drops for different values of the regularization factor $\lambda$ and the initialization variance $\sigma_\text{post}^2$. In both cases, wrong choices of the hyperparameters can hinder the correct optimization of the latent distributions. Furthermore, we tested different backbone architectures, indicating a limited performance with a VGG16 backbone. Overall the performance drops are minor and for all other settings, the model shows highly accurate results of $\kappa>0.75$.

\subsection{Required Resources} \label{sec:required_resources}
For the Gleason 2019 dataset with images of $1024\times1024$, the model can be trained with a batch size of $3$ on a single NVIDIA GeForce RTX 3090 with 24Gb memory. The training takes less than $1.5$h in total and test predictions less than $0.2$s per image. The trained model occupies less than 350Mb when saved to the disk. As each additional annotator adds only one additional vector $\mu^r\in \mathbb{R}^8$ and one covariance matrix $\Sigma^r\in \mathbb{R}^{8\times 8}$, it is scalable to a large number of annotators. The model's quick runtime and excellent scalability make it easily applicable in clinical practice.

\subsection{Limitations} \label{sec:limitations}
As semantic segmentation itself is a challenging task, some details of the annotator segmentations are not captured well by the model, such as the variations of class NC (green) in the GT of Fig. \ref{fig:inter_d} - \ref{fig:inter_x} or class G4 (orange) in the GT of Fig. \ref{fig:inter_f} - \ref{fig:inter_z}. Here, the model tends to predict similar shapes for the raters. A possible solution is to use more layers in the segmentation head $f_\theta$ with a wider kernel (e.g. $5\times5$ convolutions). This would increase the complexity of the model and might enable it to capture the different labeling behavior in even more detail. 

\begin{table}
\small
\begin{center}
\begin{tabular}{l c   l    l}
\toprule
 Hyperp. & Value     &  Unweighted $\kappa$ & Accuracy \\
\midrule
D  & $4$ &  $0.752 \pm 0.005$  & $0.836 \pm 0.003$  \\
  & $\mathbf{8}$ &  $0.758 \pm 0.011$  & $0.84 \pm 0.007$ \\
  & $16$ &  $0.752 \pm 0.006$  & $0.836 \pm 0.004$  \\
\midrule
$\sigma_\text{prior}^2$  & $1$ &  $0.758 \pm 0.007$  & $0.839 \pm 0.004$  \\
  & $\mathbf{2}$ &  $0.758 \pm 0.011$  & $0.84 \pm 0.007$  \\
  & $4$ &  $0.757 \pm 0.007$  & $0.839 \pm 0.004$  \\
\midrule
$\sigma_\text{post}^2$  & $4$ &  $0.757 \pm 0.007$  & $0.839 \pm 0.004$  \\
  & $\mathbf{8}$ &  $0.758 \pm 0.011$  & $0.84 \pm 0.007$  \\
  & $16$ &  $0.745 \pm 0.009$  & $0.83 \pm 0.005$  \\
\midrule
$\lambda$  & $0.0001$ &  $0.744 \pm 0.003$  & $0.829 \pm 0.003$  \\
  & $\mathbf{0.0005}$ &  $0.758 \pm 0.011$  & $0.84 \pm 0.007$ \\
  & $0.001$ &  $0.745 \pm 0.008$  & $0.837 \pm 0.005$  \\
\midrule
$\nu$  & $0.01$ &  $0.757 \pm 0.004$  & $0.839 \pm 0.002$  \\
  & $\mathbf{0.02}$ &  $0.758 \pm 0.011$  & $0.84 \pm 0.007$  \\
  & $0.04$ &  $0.753 \pm 0.01$  & $0.836 \pm 0.005$  \\
\midrule
Backbone & VGG16  &  $0.734 \pm 0.01$  & $0.823 \pm 0.005$  \\
  & \textbf{Resnet34}  &  $0.758 \pm 0.011$  & $0.84 \pm 0.007$  \\
  &Eff.netB2 &  $0.754 \pm 0.01$  & $0.836 \pm 0.004$  \\
\bottomrule
\end{tabular}
\end{center}
\caption{Study of hyperparameter robustness using the Gleason 2019 dataset. The default hyperparameter value is marked with bold letters. While varying one hyperparameter, all other values are set to the default value. We observe consistent and robust performance across all settings of tested hyperparameters.}
\label{tab:ablation_studies}
\end{table}

\section{Conclusions}
In this work we present “Pionono", a method for medical image segmentation that models the inter- and intra-observer variability explicitly with a probabilistic approximation. This is especially relevant for tasks where the labeling behavior of medical experts is known to vary widely, such as in the case of prostate cancer segmentation.
Our experiments on real-world cancer segmentation data demonstrate that Pionono outperforms state-of-the-art models such as STAPLE, Probabilistic U-Net, and models based on confusion matrices. Apart from the improved predictive performance, it provides a probabilistic uncertainty estimation and the simulation of expert opinions for a given test image. This makes it a powerful tool for medical image analysis and has the potential to improve the diagnostic process considerably. 

{\small
\bibliographystyle{IEEEtran}
\bibliography{ms}
}

\end{document}